\documentclass{article}
\usepackage{spconfa4,amsmath,graphicx,float,subfig}

\title{Multi-View Networks for Denoising of Arbitrary Numbers of Channels}
\twoauthors
 {Jonah Casebeer*, Brian Luc* \thanks{*These two authors contributed equally.}}
	{University of Illinois at Urbana-Champaign\\
	jonahmc2, luc2@illinois.edu}
 {Paris Smaragdis \thanks{This work was supported by NSF grant \#1319708}}
	{University of Illinois at Urbana-Champaign\\
	Adobe Research}

\begin{document}

\maketitle
\begin{abstract}
We propose a set of denoising neural networks capable of operating on an arbitrary number of channels at runtime, irrespective of how many channels they were trained on. We coin the proposed models \textit{multi-view networks} since they operate using multiple views of the same data. We explore two such architectures and show how they outperform traditional denoising models in multi-channel scenarios. Additionally, we demonstrate how multi-view networks can leverage information provided by additional recordings to make better predictions, and how they are able to generalize to a number of recordings not seen in training.
\end{abstract}
\begin{keywords}
multichannel, denoising, deep learning
\end{keywords}
\section{Introduction}
\label{sec:intro}
Suppose you are provided with multiple noisy recordings of the same event and wish to produce a single clean recording. Historically, this problem has been addressed using microphone array techniques, which can seamlessly scale to an arbitrary number of channels if the necessary hardware is in place. Although such techniques work well, recently we have seen learning-based methods being used for such tasks because of their ability to resolve much more complex denoising problems. More specifically, we have seen a strong wave of deep learning-based methods that introduce powerful non-linear models that can include more parameters and that can learn to resolve more challenging denoising problems.

Deep learning-based denoising and source separation have been explored in a variety of settings. Liu \textit{et al.} \cite{liu2014experiments} has studied deep learning for single channel denoising through spectral masking and regression  as applied independently over each spectral frame of the noisy input. Representations capable of leveraging the temporal nature of audio were introduced by Huang \textit{et al.} in 2014 and 2015 \cite{huang2014deep, huang2015joint}. In this work the authors constructed Recurrent Neural Networks (RNNs) to leverage the strong dependency between consecutive spectral frames in single-channel audio recordings. Others have since employed similar techniques \cite{weninger2014discriminatively, chien2017variational, chandna2017monoaural, qian2017speech, osako2017supervised, xu2015regression}.


Multi-channel and deep learning techniques have been combined by Swietojanski \textit{et al.}\cite{swietojanski2013hybrid} in 2013 and Araki \textit{et al.}\cite{araki2015exploring} in 2015  who both constructed multi-channel features for speech enhancement in the context of ASR. The authors of these works found that some multi-channel features could outperform conventional methods. Similarly Nugraha constructed a multi-channel framework using DNNs for estimation of the spectra of sources and an EM algorithm to combine these in to a multi-channel filter \cite{nugraha2016multichannel}. When several recordings are available these methods are able to capture new information. Li \textit{et al.} and Xiao \textit{et al.} both in 2016 where able to further leverage multiple channels with a time domain and frequency domain neural beamformers respectively \cite{li2016neural, xiao2016deep}.Now, with deep clustering Wang \textit{et al.} \cite{wang2018multi} performed multi-channel speaker independent source separation.

A common problem with learning-based methods (whether based on deep learning or not), is that the training setup needs to be replicated during inference. That means that when one trains a multi-channel system to perform, e.g., 4-channel denoising, that system can only be straightforwardly deployed on a 4-channel system. That is in contrast to analytical approaches, like classical array methods, that make use of geometric information to perform their task. The price to pay for more using more powerful deep learning models is that this analytical flexibility is lost, and one can only train and deploy using very similar setups. Here, we address this problem by introducing two neural network architectures that can be trained on a different number of channels than the number of channels that they are deployed for. This allows us to train systems on, e.g., 4-channels and deploy them on an 8-channel (or 2-channel) array without having to retrain, or in any way modify the learned models.

With evidence of RNNs surpassing feedforward networks, we propose two RNN architectures to extend current multi-channel techniques. Our first solution is to construct an RNN that instead of unrolling over time, it unrolls across the number of input channels. This allows us to train and deploy this model on an arbitrary number of input channels. We subsequently propose an additional model which unrolls both across time and channels. We call these multi-view networks (MVN), because they combine multiple views of the input. We find that they are able to consistently leverage information provided from additional recordings, as well as to generalize to a number of recordings not seen in training. 
\begin{figure}[H]
    \centering
    \includegraphics[scale=0.30]{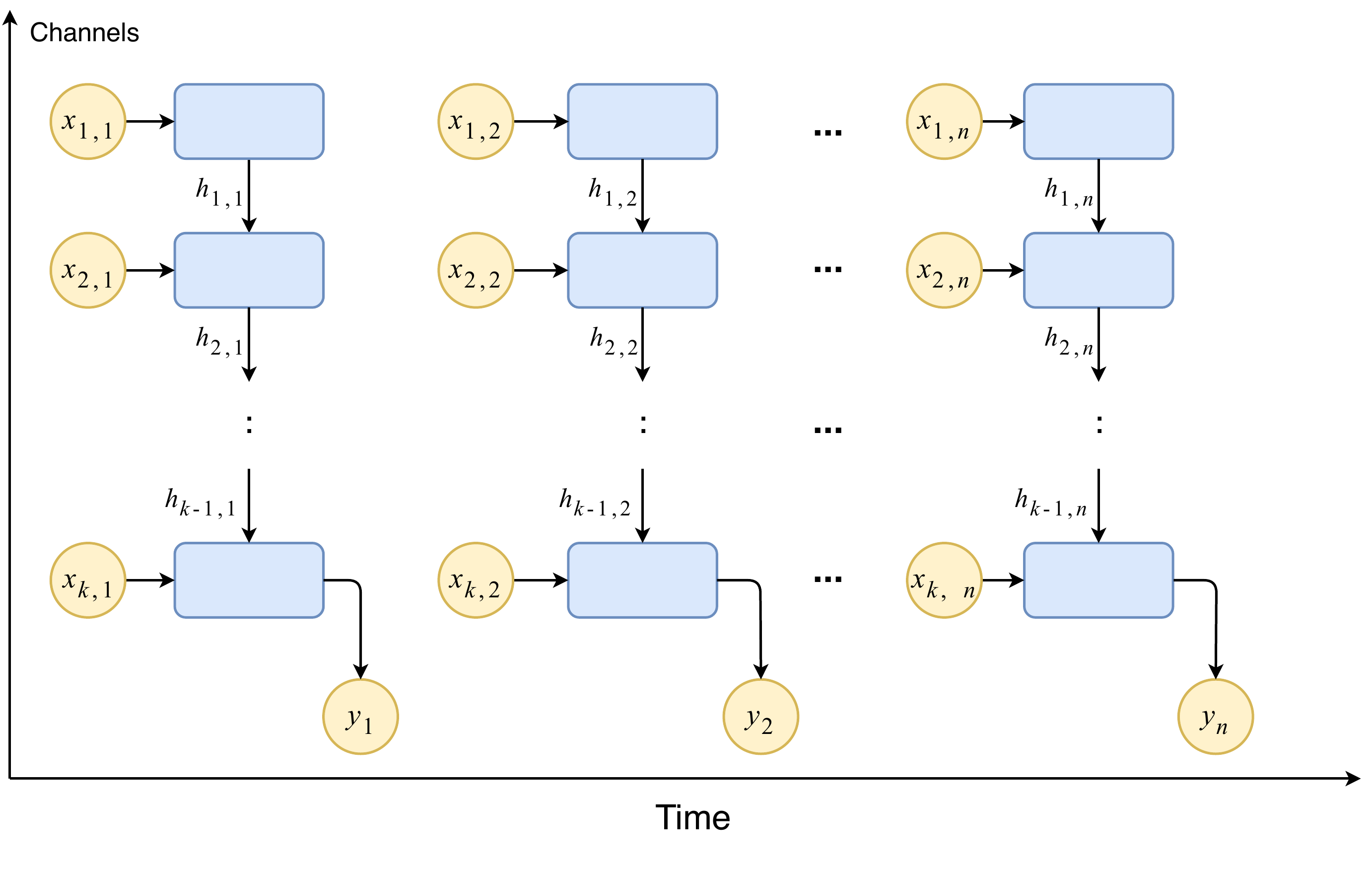}
    \setlength{\belowcaptionskip}{-10pt}
    \caption{1D MVN unrolling across channels. $x_{i, j}$ represents the $j$th spectral frame of the $i$th recording. $h_{i,j}$ represents the hidden state produced by the RNN at the $j$th spectral frame of the $i$th recording. $y_j$ is the predicted clean spectral frame.}
    \label{fig:1dunroll}
\end{figure}
\section{Multi-View Models for Denoising}
\label{sec:pagestyle}
Suppose you are provided with a noisy signal $s(t)$ which you wish to denoise. In the classic denoising RNN setup we apply a Short-Time Fourier Transform (STFT) on $s(t)$ to obtain a series of spectral magnitude frames $x_i$. An RNN would then unroll through these frames over time as follows:
\begin{align}
\begin{split}
    h_i &=\sigma(W_h x_i +U_hh_{i - 1})\\
    y_{i} &=\sigma(W_x h_{i})
    \end{split}
    \label{rnneq}
\end{align}
and would find optimal matrices $W_h$ and $U_h$ to provide a set of denoised magnitude STFT frames $y_i$. The function $\sigma$ can be any appropriate neural network activation. The unrolling scheme in the equations above takes advantage of temporal information that lies across spectral frames, and can also let us process inputs with an arbitrary length irrespective of the training data. Suppose now that you are provided with multiple noisy recordings $s_{1:k}(t)$ of the same event and wish to produce a single clean recording. We define $x_{i,j}$ to represent the $j^{th}$ spectral frame in the $i^{th}$ recording's STFT. If we wanted to use the aforementioned model and take advantage of the multiple channels, we could apply it on the averaged channel spectra, or on each input channel separately and then average all the outputs. Although this makes use of the multiple channels, these approaches are not particularly effective.

We propose using the RNN unrolling scheme across input recordings $x_{1:k,t}$ at every spectral frame $t$. This approach can allow us to take advantage of multi-channel information at each time frame, and has a number of advantages over averaging. For example, it is possible that at different points in time a different channel might provide the best input for denoising; unrolling in this fashion allows the model to leverage that instead of averaging the result with worse channels.
\begin{figure}[H]
    \centering
    \includegraphics[scale=0.30]{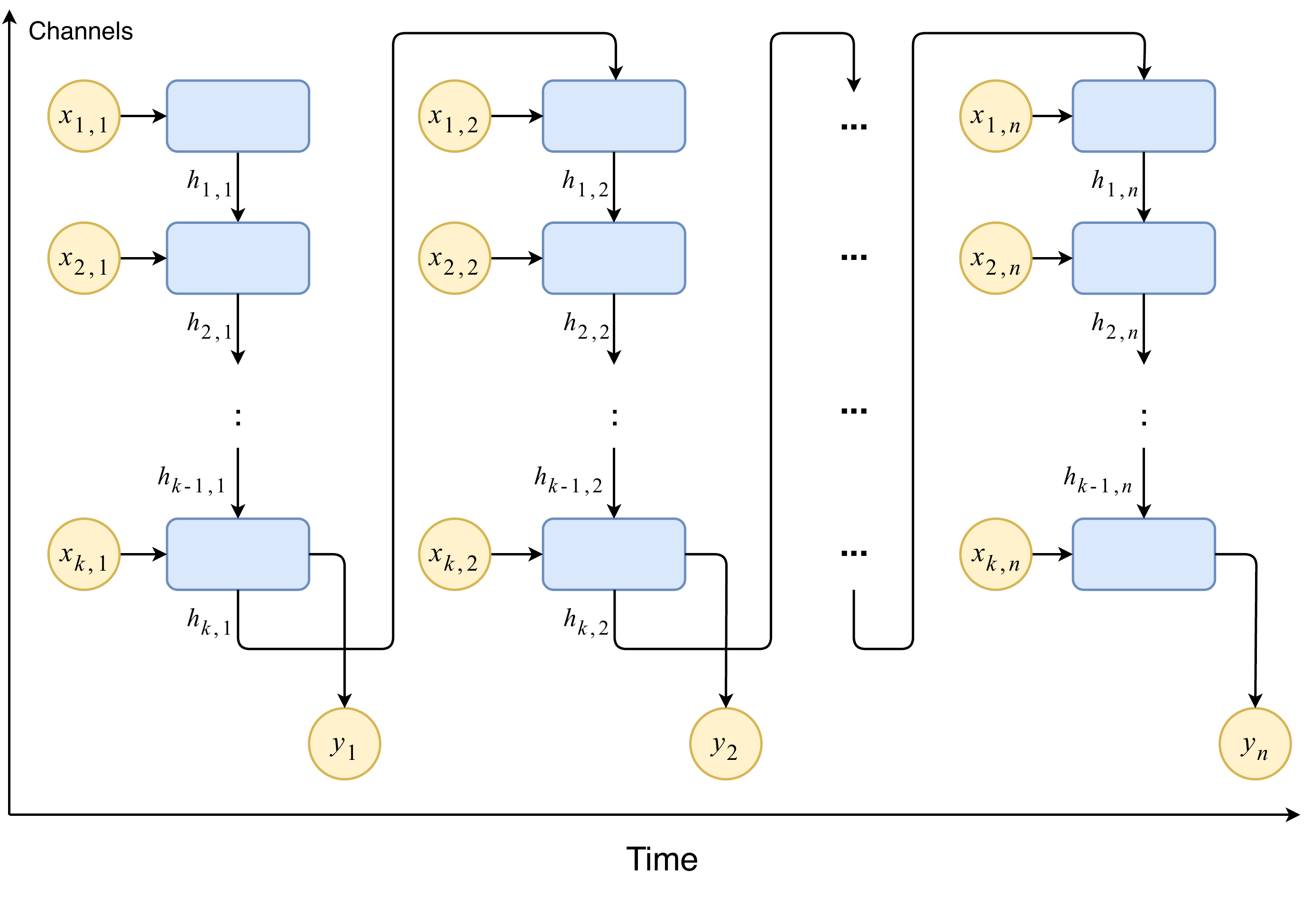}
    \caption{2D MVN unrolling across both channels and time, using the same notation as Fig. \ref{fig:1dunroll}, note how the last channel's hidden state feeds into the first channel of the next time step}
    \label{fig:2dunroll}
\end{figure}
Additionally, this approach allows us to test on an arbitrary number of channels regardless of how many we used in training. To reconstruct the denoised spectra, we experimented with averaging each output of the RNN as it unrolls over channels, as well as taking the output after the last channel is processed. We found using the last hidden state as the base for a prediction to work best. Figure \ref{fig:1dunroll} demonstrates how unrolling across channels works.
The obvious disadvantage of this approach is that we do not make use of temporal structure anymore. To address this problem we introduce a 2-dimensional RNN that unrolls across both time and sources. This allows a model to leverage the temporal dependency between time steps as well as the mutual information between different channels. Now, if the source of noise or clean signal moves with respect to the microphones the model can find the best recording to denoise and leverage previous spectral information about the sound. We accomplish this with the recurrence shown below:
\begin{align}
\begin{split}
    h_{i, j} &= 
        \begin{cases}
            \sigma(W_h x_{i,j} + U_h h_{k,j-1}) & \text{if } i=1 \\
            \sigma(W_h x_{i,j} + U_h h_{i-1,j}) & \text{otherwise}
        \end{cases}\\
    y_{j} &= \sigma(W_x h_{k,j})
\end{split}
\label{rnn2eq}
\end{align}
Note that in the case of a single input recording the 2D MVN simply unrolls across time. Figure \ref{fig:2dunroll} illustrates the 2D unrolling over channels and time.

Given these two different unrolling schemes we construct two different networks. First, for the 1D case. A denoising MVN is composed of a fully-connected front layer, a recurrent layer, and a fully-connected back layer. The front layer is given the magnitude STFT of the input channels, the recurrent layers Eq.~\eqref{rnneq} performs the unrolling operations across recordings, and the back layer regresses the RNNs output into the original STFT dimensions. To transform back to the time domain we use the phase STFT of the last channel.
\begin{figure}[h]
    \includegraphics[scale=0.12]{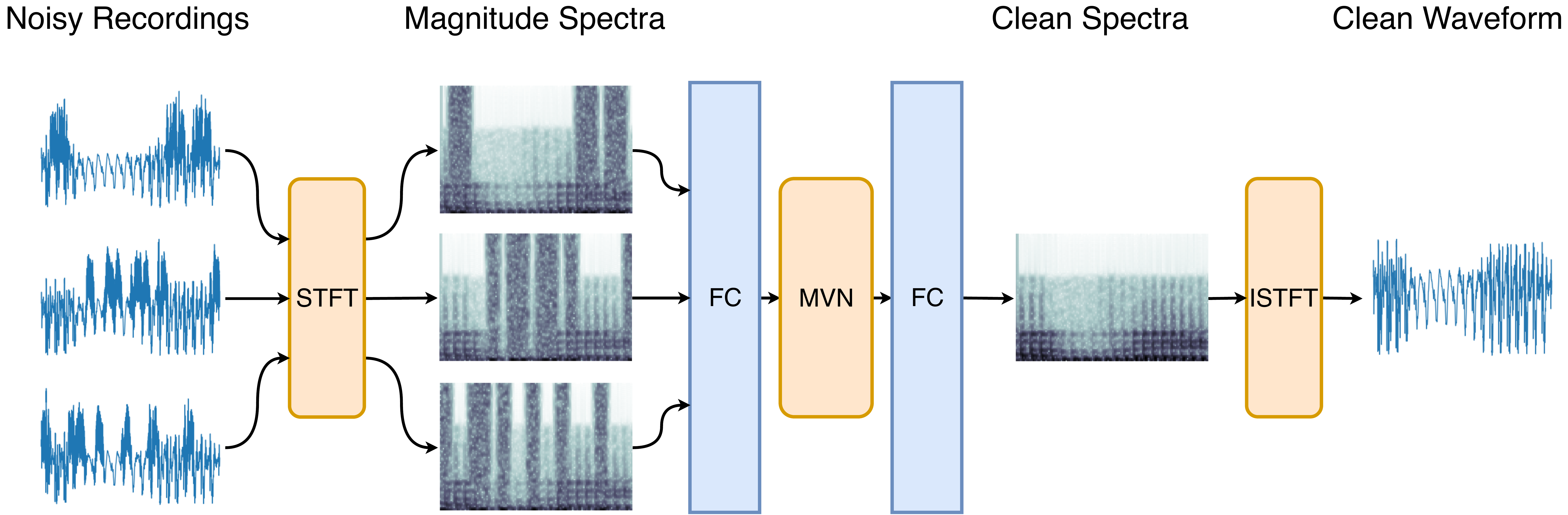}
    \setlength{\belowcaptionskip}{-10pt}
    \caption{Denoising pipeline using an MVN.}
    \label{fig:blockdiagram}
\end{figure}
For the 2D case, a denoising MVN is identical to a 1D MVN except for the recurrence which is defined in Eq.~\eqref{rnn2eq}. Figure \ref{fig:blockdiagram} broadly illustrates an MVN supplied with 3 noisy recordings containing mutual information. In practice, we used GRUs \cite{cho2014learning} instead of plain RNNs.

We find that training MVNs with an SDR-proxy loss as defined in \cite{venkataramani2017adaptive} works far better than traditional norm-based loss functions. The loss used is:
\begin{equation}
    SDRLoss(x, y) = -\frac{ (x^\top y)^2 }{x^\top x}
\end{equation}
where $y$ and $x$ are vectors containing the target and output time-domain signals respectively.

\section{Experiments}
\label{sec:typestyle}
Our system was evaluated on two different kinds of noisy mixtures. Both are created from speakers in the TIMIT data set and from segments of "Babble", "Airport", "Train", and "Subway" noises. From TIMIT we select 12 female speakers each of which has 10 unique utterances \cite{garofolo1993darpa}. From the 120 utterances, we randomly select 100 for training and 20 for validation. From each utterance we create a two-second noisy mixture by adding one of the "Babble", "Airport", "Train" and "Subway" noises. We propose two mixing techniques and evaluate our performance with the BSS-Eval Metric: Source to Distortion Ratio (SDR) \cite{fevotte2005bss_eval}. In both setups we only show results for the 2D MVN as it outperforms the 1D MVN.
\subsection{Naive Averaging RNN}
\label{ssec:subhead}
As a benchmark we used an averaging model. This model averages all channels then passes the averaged STFT frames through a dense layer which reduces 1024-point DFTs to a size of 512. We then unroll a GRU with a hidden size of 512. The hidden size is expanded from 512 dimensions to 1024 with a dense layer. Finally, we perform an ISTFT to produce a denoised signal. We refer to this model as the averaging RNN. This model is trained with the softplus nonlinearity.
\begin{figure}[h]
    \centering
    \includegraphics[scale=0.62]{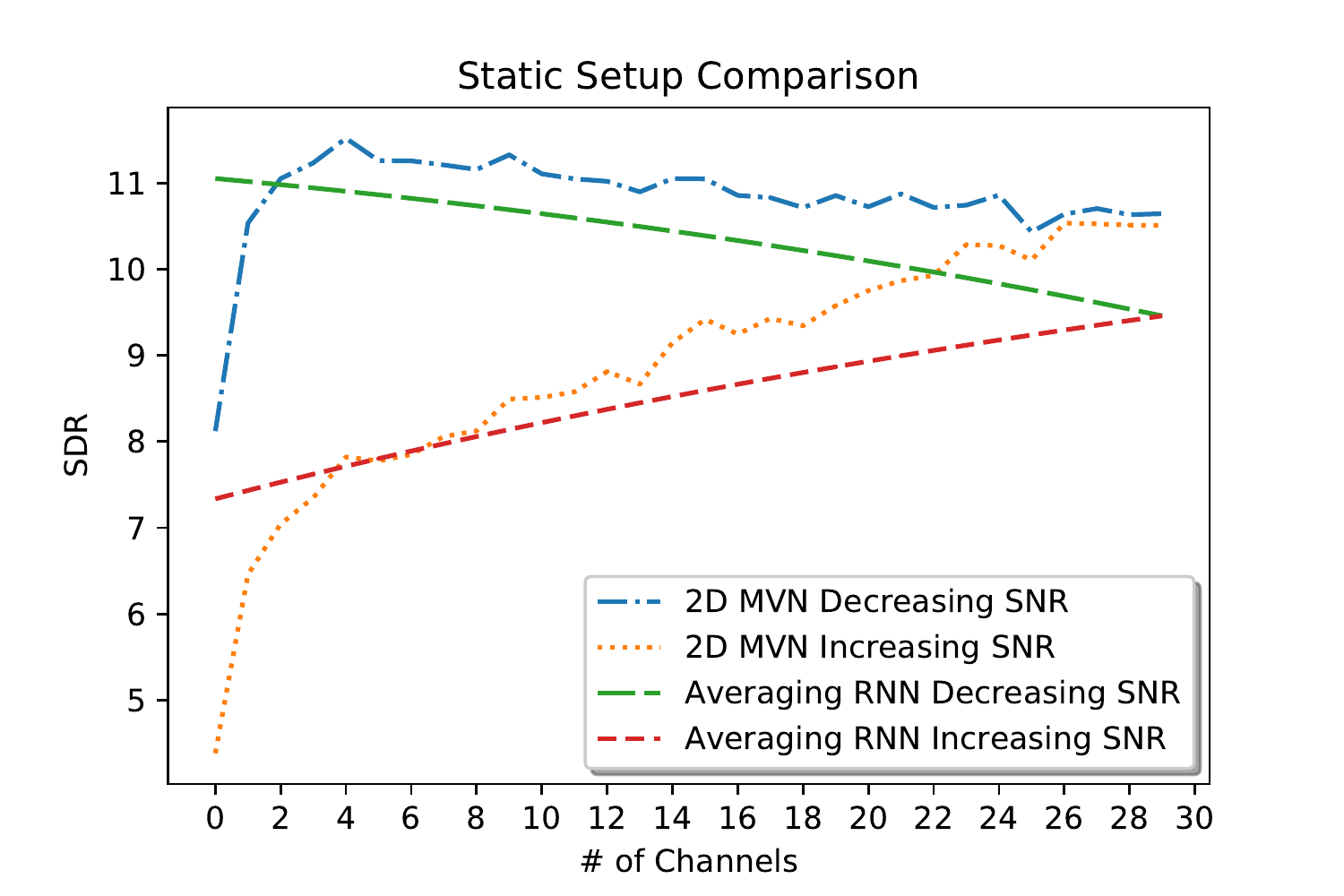}
    \setlength{\belowcaptionskip}{-10pt}
    \caption{Testing how 2D MVNs leverage new channel information. The x-axis on the plot denotes the number of provided noisy channels, and the y-axis shows the average SDR on the validation set. In the case of decreasing SNR the MVN obtains peak performance after a few channels and then largely retains it, not being influenced by the latter poor-quality input channels. With increasing SNR, the MVN keeps performing better as additional cleaner channels are being received. Ultimately, the MVN performs the same on both cases (which both exhibit recordings from $-5$ to $5$ dB but in reverse order), showing it's performance is permutation invariant.}
    \label{fig:inc_2d}
\end{figure}
\subsection{Model Parameters}
\label{ssec:subhead}
We construct a 2D MVN which operates on 1024-point DFTs, has a fully-connected layer that goes from 1024 to 512 dimensions, and then a GRU over channels with a hidden size of 512. This hidden size is projected back to 1024 dimensions with a final fully-connected layer. All models used the softplus nonlinearity.

\subsection{Static Noise Setup}
\label{ssec:subhead}
Because RNNs process inputs serially we need to verify that the output we obtain is not mostly dependent on the last few observed channels. More specifically, we want to see that this model can leverage cleaner inputs (and ignore excessively noisy inputs) regardless of whether they are presented first or last. And more generally, we want to ensure that performance is invariant to input presentation order. In order to do that, we simulate a static mixing scenario with a static target and noise, and randomly placed static microphones.

In this setup we examine two cases. In the first case we use $k$ randomly-ordered input channels which exhibit an SNR ranging from $-5$ to $-5 + k/3$ dB, with $k \in [1,\dots,30]$. That way for, e.g., six input channels we would observe SNRs ranging from $-5$ to $-3$ dB. Conversely, in the second case the input SNRs for $k$ channels are ordered from $5$ to $5-k/3$ dB. In both cases, once $k=30$ we will see recordings varying from $-5$ to $5$ dB, but for $k<30$ depending on the case we will either see a set of cleaner recordings, or one of noisier recordings. The goal here is to see how this approach gets swayed by the presentation order of the input channels.

For both increasing and decreasing SNR scenarios, we test a 2D MVN trained on five channels with random order SNRs. The results are shown in Figure \ref{fig:inc_2d}. For reference we compare the 2D MVN with the baseline RNN.

We observe that MVNs outperform the averaging RNN at leveraging new information and that they are good at ignoring noisy channels and not being swayed by the input channel ordering. This is a very desirable behavior as it means we can safely provide this denoiser with multiple recordings, without having to worry about their ordering and whether the best inputs come first or last.

\subsection{Dynamic Noise Setup}
\label{ssec:subhead}
\begin{figure}[h]
    \centering
    \includegraphics[width=0.5\textwidth]{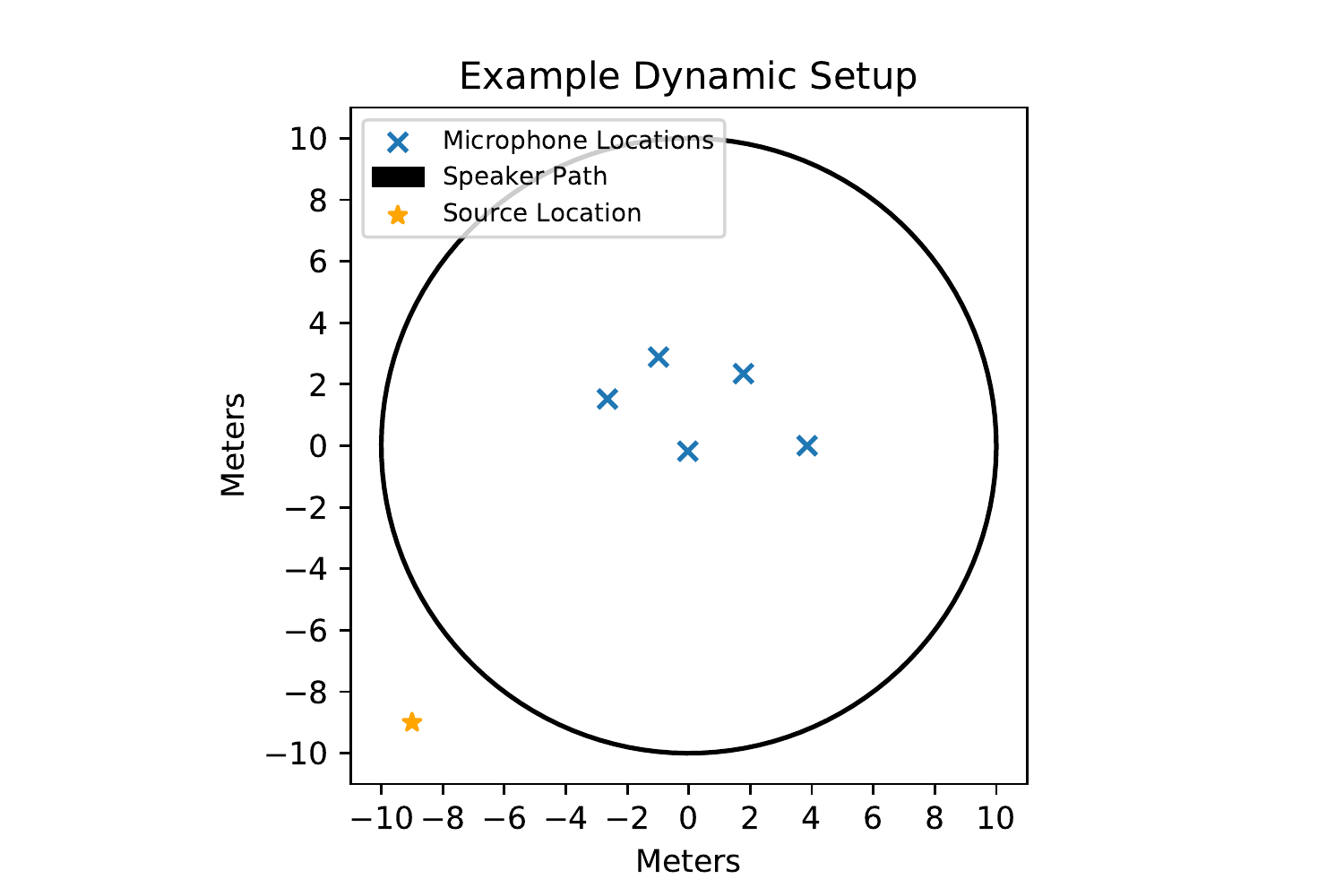}
    \setlength{\belowcaptionskip}{-10pt}
    \caption{Example dynamic noise setup with five channels. The noise source moves along a circle, changing which microphone gets the best input each time.}
    \label{fig:dynamicsetup}
\end{figure}

The second setup replicates a physically stationary target source, a noise source moving in a circle, and stationary microphones randomly placed within the circle formed by the noisy source path. We set the average SNR across each mixture to $0$ while the instantaneous SNR is dependent on the microphone-source geometry. Figure \ref{fig:dynamicsetup} illustrates this setup. Here we reward models capable of leveraging the dynamic nature of the instantaneous SNR for each recording. We call this the dynamic setup.

In training we generated samples with five channels. Every training sample simulates a new setup where mic placement and ordering has been randomized. This means our model can't memorize any particular mic setup. However, at test time we maintain the same setup for each different number of microphones to more accurately show how providing additional noisy channels to our model changes performance.

Using this setup we generate a single result plot. The x-axis denotes the number of noisy channels provided to the model. The y-axis shows the average SDR in the validation set for that number of channels. The horizontal line across the graph denotes the performance of the averaging RNN model, which mostly stays constant regardless of the input channels.
\begin{figure}[h]
    \centering
    \includegraphics[scale=0.63]{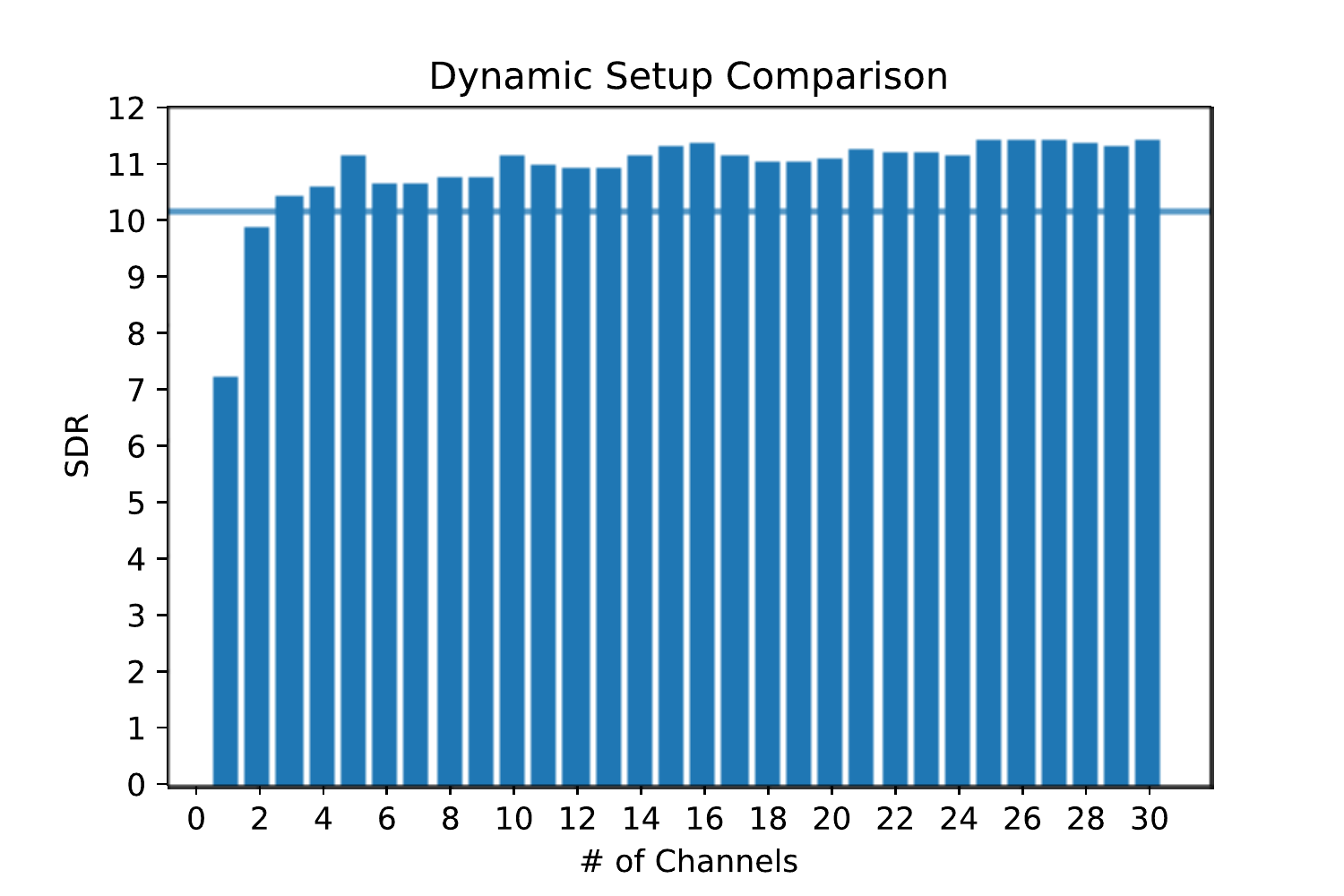}
    \setlength{\belowcaptionskip}{-10pt}
    \caption{2D MVN performance vs number of recordings. Even though the network was trained on 5 channels (note the peak), the quality of outputs keeps improving past 5 channels. In contrast, the regular averaging RNN cannot take advantage of additional channel information. This plot corresponds to a bidirectional GRU across channels.}
    \label{fig:fake2dDynamicplot}
\end{figure}
We observe that the 2D MVN outperforms the averaging RNN and is able to leverage numbers of channels far beyond the amount it was trained on. It does not however do as well when only observing one or two channels. We hypothesize that this is because it doesn't fully utilize the recurrent connection in these cases. Furthermore, the MVN can do so even when the "best" recording is different at every time step. This ability is exciting since new information with each channel can be always leveraged, without necessitating changes in the model.
     
\section{Conclusion}
\label{sec:majhead}
We have proposed a denoising RNN capable of operating on an arbitrary number of input recordings and leveraging new channel information. We show how the order of the channels does not influence the quality of the results, and that its denoising ability keeps improving as we provide more input channels, even past the amount we trained on. Finally we show how this network outperforms the alternative approach of averaging the input channels. Although not shown here due to space constraints, this model can also operate on an arbitrary number of recordings at every time step, allowing for deployment on settings with a dynamically changing number of sensors. 

\bibliographystyle{IEEEbib}
\bibliography{refs}
\end{document}